# Impact of graphene polycrystallinity on the performance of graphene field-effect transistors

David Jiménez[1], Aron W. Cummings[2], Ferney Chaves[1], Dinh Van Tuan[2], Jani Kotakoski[3,4], Stephan Roche[2,5]

[1]Departament d'Enginyeria Electrònica, Escola d'Enginyeria,
Universitat Autònoma de Barcelona, 08193-Bellaterra, Spain

[2]ICN2, Institut Català de Nanociencia i Nanotecnologia, Campus UAB, 08193 Bellaterra (Barcelona), Spain

[3]Faculty of Physics, University of Vienna, Boltzmanngasse 5, 1090 Wien, Austria

[4]Department of Physics, University of Helsinki, P.O. Box 43, 00014 University of Helsinki, Finland

[5]ICREA, Institució Catalana de Recerca i Estudis Avançats, 08070 Barcelona, Spain.

*Dated 18-Dec-2013*

**Abstract**

We have used a multi-scale physics-based model to predict how the grain size and different grain boundary morphologies of polycrystalline graphene will impact the performance metrics of graphene field-effect transistors. We show that polycrystallinity has a negative impact on the transconductance, which translates to a severe degradation of the maximum and cutoff frequencies. On the other hand, polycrystallinity has a positive impact on current saturation, and a negligible effect on the intrinsic gain. These results reveal the complex role played by graphene grain boundaries, and can be used to guide the further development and optimization of graphene-based electronic devices.

In the effort to successfully realize next-generation technologies based on graphene field-effect transistors (GFETs), theory and device modeling will play a crucial role. Specifically, it is important to develop models that can accurately describe both the electrostatics and the current-voltage (*I-V*) characteristics of graphene-based electronic devices.[1-10] This capability will enable device design optimization and performance projections, will permit benchmarking of graphene-based technology against existing ones,[11-12] and will help to explore the feasibility of analog/RF circuits based on graphene.[13-15] Ultimately, graphene-based devices could provide new or improved functionality with respect to existing technologies, such as those based on silicon or III-V materials.

The chemical vapor deposition (CVD) technique for growing wafer-scale graphene on metallic substrates[16-19] produces a polycrystalline pattern. This is because the growth of graphene is simultaneously initiated at different nucleation sites, leading to samples with randomly distributed grains of varying lattice orientations.[20] It has recently been predicted that the electronic properties of polycrystalline graphene differ from those of pristine graphene (PG), where the mobility scales linearly with the average grain size.[21] Based on these results, we report on how the electronic properties of polycrystalline graphene (Poly-G) impact the behavior of graphene-based devices. Specifically, we concentrate our study on the effect that Poly-G has on the gate electrostatics and *I-V* characteristics of GFETs. We find that the source-drain current and the transconductance are proportional to the average grain size, indicating that these quantities are hampered by the presence of grain boundaries (GBs) in the Poly-G. However, our simulations also show that current saturation is improved by the presence of GBs, and the intrinsic gain is insensitive to the grain size. These results indicate that GBs play a complex role in the behavior of graphene-based electronics, and their importance depends on the application of the device.

The starting point of our study is the characterization of a large-area model of disordered Poly-G samples, containing hundreds of thousands atoms and described by varying grain misorientation angles, realistic carbon ring statistics, and unrestricted GB structures, based on the method reported in Ref. 22. To calculate the electronic and transport properties we used a tight-binding (TB) Hamiltonian and an efficient quantum transport method,[23,24] which is particularly well-suited for large samples of disordered low-dimensional systems. The transport calculations were based on a real-space order-N quantum wave packet evolution approach, which allowed us to compute the Kubo-Greenwood conductivity $\sigma(E) = \frac{e^2}{4} DOS_{p-G}(E) \lim_{t \to \infty} \frac{\partial \Delta X^2(E,t)}{\partial t}$, where $DOS_{p-G}(E)$ is the density of states of the Poly-G and $\Delta X^2(E,t)$ is the mean-square spreading of the wave packet. With this quantity, the charge carrier mobility can be estimated as µ(E)=σ(E)/q*$Q_c$(E), where $Q_C$ is the 2D charge density in the graphene. It should be noted that we assume the carrier mobility is not limited by the substrate, that is, we do not consider additional scattering due to charge traps or surface phonons in the insulator that could further degrade the carrier mobility[25]. Thus, our results represent an upper bound on the performance metrics of the GFETs that we are studying.

In this work we focus on a dual-gate GFET as the one depicted in Fig. 1. This transistor is based on a metal/oxide/Poly-G/oxide/semiconductor structure where an external electric field modulates the mobile carrier density in the Poly-G layer. The electrostatics of this dual gate structure can be understood with an application of Gauss's law,

$$Q_c = C_t\left(V_{gs}^* - V_c\right) + C_b\left(V_{bs}^* - V_c\right), \qquad (1)$$

where $Q_c=q(p-n)$ is the net mobile charge density in the graphene channel, $C_t$ and $C_b$ are the geometrical top and bottom oxide capacitances, and $V_{gs}^*$ and $V_{bs}^*$ are the effective top and bottom gate-source voltages, respectively. Here, $V_{gs}^*=V_{gs}-V_{gs0}$ and $V_{bs}^*=V_{bs}-V_{bs0}$, where $V_{gs0}$ and $V_{bs0}$ are quantities that comprise the work function differences between each gate and the graphene channel, charged interface states at the graphene/oxide interfaces, and possible doping of the graphene. The graphene charge density can be determined numerically using the procedure

$$Q_c(V_c) = q\int_{-\infty}^{0} DOS_{p-G}(E)f(qV_c - E)dE - q\int_{0}^{\infty} DOS_{p-G}(E)f(E - qV_c)dE, \qquad (2)$$

where $DOS_{p-G}(E)$ has been calculated with the procedure outlined in Ref. 21. The potential $V_c$ represents the voltage drop across the graphene layer, and is related to the quantum capacitance $C_q$ of the Poly-G by $C_q = -dQ_c/dV_c$. When the entire length of the transistor is considered, the effective gate voltages can be written as $V_{gs}^*=V_{gs}-V_{gs0}-V(x)$ and $V_{bs}^*=V_{bs}-V_{bs0}-V(x)$, where $V(x)$ (the so-called quasi-Fermi level) represents the potential along the graphene channel. The boundary conditions that should be satisfied are $V(0)=0$ at the source and $V(L)=V_{ds}$ at the drain.

To model the drain current, we employ a drift-diffusion model with the form $I_{ds} = -W|Q_c(x)|v(x)$, where W is the gate width, $Q_c(x)$ is the free carrier sheet density in the channel at position x, and $v(x)$ is the carrier drift velocity. The latter is related to the transverse electric field E as $v=\mu E$, so no velocity saturation effect has been included in this model. The low-field carrier mobility $\mu(Q_c)$ is density-dependent and calculated via the procedure of Ref. 21. After applying E=-dV(x)/dx, including the above expression for v, and integrating the resulting equation over the device length, the source-drain current becomes

$$I_{ds} = \frac{W}{L} \int_0^{V_{ds}} \mu |Q_c| dV. \qquad (3)$$

In order to calculate $I_{ds}$, the integral in (3) is solved using $V_c$ as the integration variable and subsequently expressing $\mu$ and $Q_c$ as functions of $V_c$, based on the mapping given by (2). This gives

$$I_{ds} = \frac{W}{L} \int_{V_{cs}}^{V_{cd}} \mu(V_c) |Q_c(V_c)| \frac{dV}{dV_c} dV_c, \qquad (4)$$

where $V_c$ is obtained by self-consistently solving (1) and (2). The channel potential at the source is determined as $V_{cs}=V_c(V=0)$ and the channel potential at the drain is determined as $V_{cd}=V_c(V=V_{ds})$. Finally, (1) allows us to evaluate the derivative appearing in (4), namely $\frac{dV}{dV_c} = -1 + \frac{C_q}{C_t + C_b}$, which should be determined numerically as a function of the integration variable $V_c$.

Next, we apply the multi-scale model to the GFET shown in Fig. 1. It consists of a dual-gate structure with L=10 μm and W=5 μm. The top and bottom gate insulators are hafnium oxide and silicon oxide with thicknesses of 4 nm and 300 nm, respectively. For the active channel, we considered poly-G with different average grain sizes together with the simple PG case, which serves as a convenient reference for comparison. For this study, we created samples with three different average grain sizes (average diameter <d> ≈ 13, 18, and 25.5 nm) and uniform grain size distributions. The atomic structure at the GBs consists predominantly of five- and seven-member carbon rings and assumes meandering shapes similar to the experimentally observed ones. We also created one sample with <d> ≈ 18 nm and "broken" (poorly connected) boundaries ("br-18 nm"). The quantum capacitance ($C_q$) of each sample is presented in Fig. 2a, which reflects the structure of the DOS, shown in Fig. 2b. An enhanced density of zero-energy modes around the charge neutrality point (CNP) can be observed, which arises locally from the atomic configurations of the GBs, giving rise to a finite $C_q$. A zero $C_q$ would correspond to ideal gate efficiency, meaning that the gate voltage would have 100% control over the position of the graphene Fermi level. Away from the CNP, both $C_q$ and the DOS of the analyzed structures look very similar. For the poorly connected sample "br-18 nm" a peak is observed around the CNP because of a higher density of midgap states, resulting in a negative differential $C_q$.

Fig. 3a shows the transfer characteristics of the GFET under consideration for different grain sizes. The low-field carrier mobility was calculated from the Kubo-Greenwood conductivity as $\mu(E)=\sigma(E)/q*Q_c(E)$, and has been plotted as a function of $Q_c$ in Fig. 3b. The mobility corresponding to a grain size of 1 μm was estimated from the mobility at 25.5 nm with a simple scaling law,[21] $\mu_{1\mu m}(Q_c) = (1 \mu m / 25.5 nm) * \mu_{25.5nm}(Q_c)$. The resulting *I-V* characteristics exhibit the expected V-like shape with an ON-OFF current ratio in the range of 2 to 4, and one can see that the source-drain current is proportional to the average grain size. This is due to the scaling of the mobility with grain size, as shown in Fig. 3b. In Fig. 3c we plot the transconductance of the GFET, defined as $g_m=dI_{ds}/dV_{gs}$, which is a key parameter in determining the transistor voltage gain or the maximum operation frequency. It appears that small grain sizes are detrimental to this factor. The reason behind such a degradation is the combination of two factors as the grain size is reduced: (a) an increase of $C_q$ at low carrier densities (Fig. 2a), which is related with the increase of the DOS near the CNP (Fig. 2b) and leads to reduced gate efficiency; and (b) the reduction of the low-field carrier mobility (Fig. 3b) because of scattering due to the disordered atomic structure of the GBs. Fig. 3b indicates that the mobility is proportional to the average grain size of the Poly-G; a higher density of GBs results in more scattering and a lower mobility. The scattering effect of the GBs has been further quantified in Ref. 21, which shows the scaling of the conductivity and the mean free path of the Poly-G for different grain sizes. For example, the sample with 25.5-nm grains has a mean free path of 10 nm near the Dirac point, compared to 5 nm for the sample with 13-nm grains.

In Fig. 4a we plot the GFET output characteristics for different grain sizes and gate biases. The output characteristic exhibits an initial linear region dominated by hole transport (*p*-type channel), followed by a weak saturation region. The onset of saturation ($V_{sd,sat}$) happens when the channel becomes pinched off at the drain side. A further increase of $V_{sd}$ drives the transistor towards the second linear region, characterized by a channel with a mixed *p*- and *n*-type behavior. Interestingly, a reduction of the grain size improves the current saturation, which can be seen in a plot of the output conductance (Fig. 4b), defined as $g_d = dI_{ds}/dV_{ds}$. Here, the minimum of $g_d$ is much flatter and broader for smaller grain sizes. Both $g_m$ and $g_d$ determine the intrinsic gain $A_v = g_m/g_d$, which is a key figure of merit in analog or RF applications. Our simulations demonstrate that $A_v$ is insensitive to the grain size (Fig. 5), because an increase in $g_m$ is almost exactly compensated by a similar increase in $g_d$. This suggests that polycrystallinity is not a limiting factor in analog/RF devices whose performance depends on the intrinsic gain. However, there are other performance metrics, such as the intrinsic cutoff ($f_T$) and maximum frequencies ($f_{max}$), which are severely degraded by the presence of GBs. To demonstrate this, we have calculated both $f_T$ and $f_{max}$ for the device under consideration, but assuming a channel length of 100 nm. The cutoff frequency is given by $f_T \approx g_m/2\pi C_{gs}$, where $C_{gs}$ is the gate-to-source capacitance.[12] Given that the geometrical capacitance $C_t$ is much smaller than the quantum capacitance $C_q$, $C_{gs} \cong C_t$. The maximum frequency is given by $f_{max} \approx g_m/\left(4\pi C_{gs}\sqrt{g_d(R_S + R_G)}\right)$, where $R_S$ and $R_G$ are the source and gate resistances, respectively.[12] Here we have assumed state of the art values such as[26] $R_S \sim 100$ $\Omega$-$\mu$m and $R_G \sim 6$ $\Omega$. As shown in Fig. 6, $f_{max}$ and $f_T$ are degraded by one and two orders of magnitude, respectively, when the average grain size decreases from 1 $\mu$m to 25 nm.

Realistic GFETs are limited in performance by interaction with the substrate and top gates. Comparing with the extracted mobility from some reported state-of-the-art devices[27], our calculations, which represent the limiting case of uncovered graphene, overestimate the mobility of these devices by ~10x. As a consequence, $g_m$, $g_d$ and $f_T$ should be reduced by that amount when considering substrate and top gate effects. Meanwhile, $A_v$ is expected to remain constant and $f_{max}$ is expected to be reduced by ~3x. The mentioned 10x factor of mobility reduction could be made significantly smaller by using an appropriate substrate such as diamond-like carbon[28] (DLC), which helps to minimize interaction with the substrate.

In conclusion, we have developed a drift-diffusion transport model for the GFET, based on a detailed description of electronic transport in poly-G. This model allows us to determine how a graphene sample's polycrystallinity alters the electronic transport in GFETs, enabling the prediction and optimization of various figures of merit for these devices. We have found that the presence of GBs produces a severe degradation of both the maximum frequency and the cutoff frequency, while the intrinsic gain remains insensitive to the presence of GBs. Overall, polycrystallinity is predicted to be an undesirable trait in GFETs targeting analog or RF applications.

**Acknowledgments**

We acknowledge support from SAMSUNG within the Global Innovation Program. The research leading to these results has received funding from Ministerio of Economía y Competitividad of Spain under the project TEC2012-31330 and MAT2012-33911, and from the European Union Seventh Framework Programme under grant agreement n°604391 Graphene Flagship.

**References**

[1] I. Meric, M. Y. Han, A. F. Young, B. Ozyilmaz, P. Kim, K. Shepard, *Nat. Nanotechnol.* **3**, 654-659 (2008).

[2] D. Jiménez, *Nanotechnology* **19**, 345204 (2008).

[3] S. Thiele, J. A. Schaefer, F. Schwierz, *J. Appl. Phys.* **107,** 094505 (2010).

[4] D. Jiménez, O. Moldovan, *IEEE Trans. Electron Devices* **58**, 4049-4052 (2011).

[5] S. Thiele, F. Schwierz, *J. Appl. Phys.* **110**, 034506 (2011).

[6] J. Champlain, *J. Appl. Phys.* **109**, 084515 (2011).

[7] H. Wang, A. Hsu, J. Kong, D. A. Antoniadis, T. Palacios, *IEEE Trans. Electron Devices* **58**, 1523 - 1533 (2011).

[8] D. Jiménez, *IEEE Trans. Electron Devices* **58**, 4377-4383 (2011).

[9] O. Habibpour, J. Vukusic, J. Stake, *IEEE Trans. Electron Devices* **59**, 968-975 (2012).

[10] T. Fang, A. Konar, H. Xing, D. Jena, *Appl. Phys. Lett.* **91**, 092109 (2007).

[11] F. Schwierz, *Nat. Nanotechnol.* **5**, 487-496 (2010).

[12] F. Schwierz, *Proceedings of the IEEE* **101**, 1567 - 1584 (2013).

[13] Y-M. Lin, A. Valdes-Garcia, S-J. Han, D. B. Farmer, I. Meric, Y. Sun, Y. Wu, C. Dimitrakopoulos, A. Grill, P. Avouris, K. A. Jenkins, *Science* **332**, 1294-1297 (2011).

[14] H. Wang, A. Hsu, J. Wu, J. Kong, T. Palacios, *IEEE Electron Device Lett.* **31**, 906-908 (2010).

[15] H. Wang, D. Nezich, J. Kong, T. Palacios, *IEEE Electron Device Lett.* 30, 547–549 (2009).

[16] X. S. Li, W. W. Cai, J. H. An, S. Kim, J. Nah, D. X. Yang, R. Piner, A. Velamakanni, I. Jung, E. Tutuc, S. K. Banerjee, L. Colombo, R. S. Ruoff, *Science* **324,** 1312–1314 (2009).

[17] A. Reina, X. Jia, J. Ho, D. Nezich, H. Son, V. Bulovic, M. S. Dresselhaus, J. Kong, *Nano Lett.* 9, 30 (2009).

[18] X. S. Li, C. W. Magnuson, A. Venugopal, J. H. An, J. W. Suk, B. Y. Han, M. Borysiak, W. W. Cai, A. Velamakanni, Y. W. Zhu, L. F. Fu, E. M. Vogel, E. Voelkl, L. Colombo, R. S. Ruoff, *Nano Lett.* **10**, 4328–4334 (2010).

[19] S. Bae, H. Kim, Y. Lee, X. F. Xu, J. S. Park, Y. Zheng, J. Balakrishnan, T. Lei, H. R. Kim, Y. I. Song, Y. J. Kim, K. S. Kim, B. Ozyilmaz, J. H. Ahn, B. H. Hong, S. Iijima, *Nat. Nanotechnol.* **5**, 574–578 (2010).

[20] P. Y. Huang, C. S. Ruiz-Vargas, A. M. van der Zande, W. S. Whitney, M. P. Levendorf, J. W. Kevek, S. Garg, J. S. Alden, C. J. Hustedt, Y. Zhu, J. Park, P. L. McEuen, D. A. Muller, *Nature* **469**, 389 (2011).

[21] D. V. Tuan, J. Kotakoski, T. Louvet, F. Ortmann, J. C Meyer, S. Roche, *Nano Lett.* **13**, 1730-1735 (2013).

[22] J. Kotakoski, J. C. Meyer, *Phys. Rev. B* **85**, 195447 (2012).

[23] S. Roche, *Phys. Rev. B* **59**, 2284 (1999).

[24] S. Roche, N. Leconte, F. Ortmann, A. Lherbier, D. Soriano, J-C Charlier *Solid State Communications* **152**, 1404-1410 (2012).

[25] Y. Wu, Y-M Lin, A. A. Bol, K. A. Jenkins, F. Xia, D. B. Farmer, Y. Zhu, P. Avouris, *Nature* **472**, 74-78 (2011).

[26] D. B. Farmer, A. Valdes-Garcia, C. Dimitrakopoulos, P. Avouris, *Appl. Phys. Lett.* **101**, 143503 (2012).

[27] M. C. Lemme, T. Echtermeyer, M. Baus, B. N. Szafranek, M. Schmidt, H. Kurz, *ECS Transactions* **11**, 413 (2007).

[28] Y. Wu, Y. Lin, A. A. Bol, K. A. Jenkins, F. Xia, D. B. Farmer, Y. Zhu, P. Avouris, *Nature* **472**, 74 (2011).

**Figure captions:**

**Fig. 1**. (a) Schematic of the dual-gate GFET, consisting of a poly-G channel on top of an insulator layer, which is grown on a heavily-doped Si wafer acting as the back gate. An artistic view of the patchwork of coalescing graphene grains of varying lattice orientations and size is shown in (b). The source and drain electrodes contact the poly-G channel from the top and are assumed to be ohmic. The source is grounded and considered the reference potential in the device. The electrostatic modulation of the carrier concentration in graphene is achieved via a top-gate stack consisting of the gate dielectric and the gate metal.

**Fig. 2**. Quantum capacitance (a) and density of states (b) of polycrystalline graphene considering different average grain sizes. The pristine graphene (PG) case has also been plotted for the sake of comparison.

**Fig. 3**. Transfer characteristics (a) and transconductance (c) of the graphene field-effect transistor considering different samples of polycrystalline graphene as the active channel. (b) Estimated low-field carrier mobility as a function of the carrier density for each of the samples.

**Fig. 4**. Output characteristics (a) and output conductance (b) of the graphene field-effect transistor considering different samples of polycrystalline graphene as the active channel.

**Fig. 5**. Intrinsic gain as a function of the drain voltage. The transconductance and output conductance are also plotted at $V_{gs}$=-0.25V.

**Fig. 6**. Intrinsic maximum and cutoff frequency for the simulated transistor assuming a channel length of 100 nm.

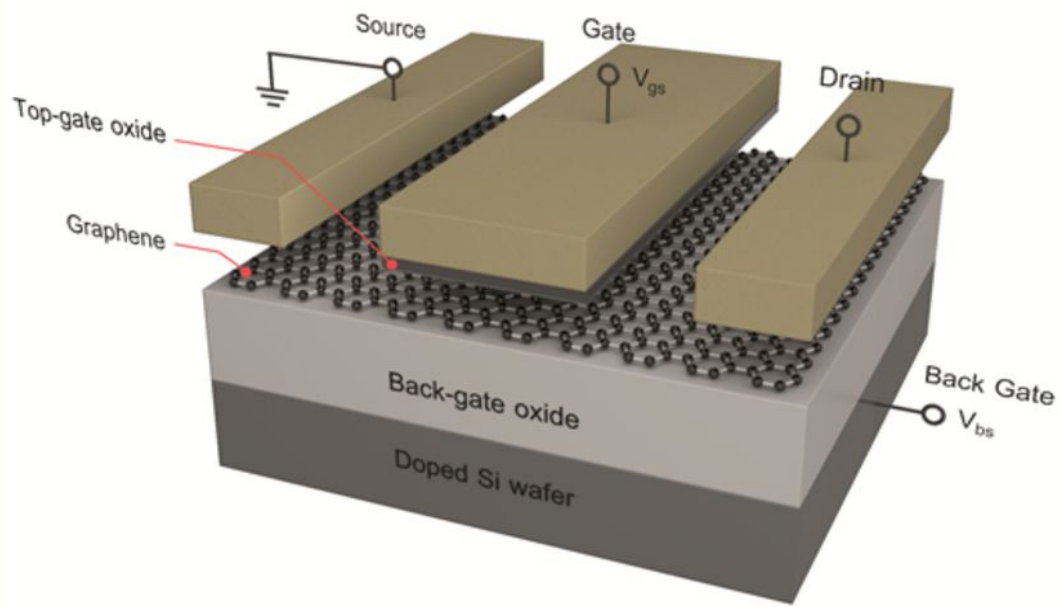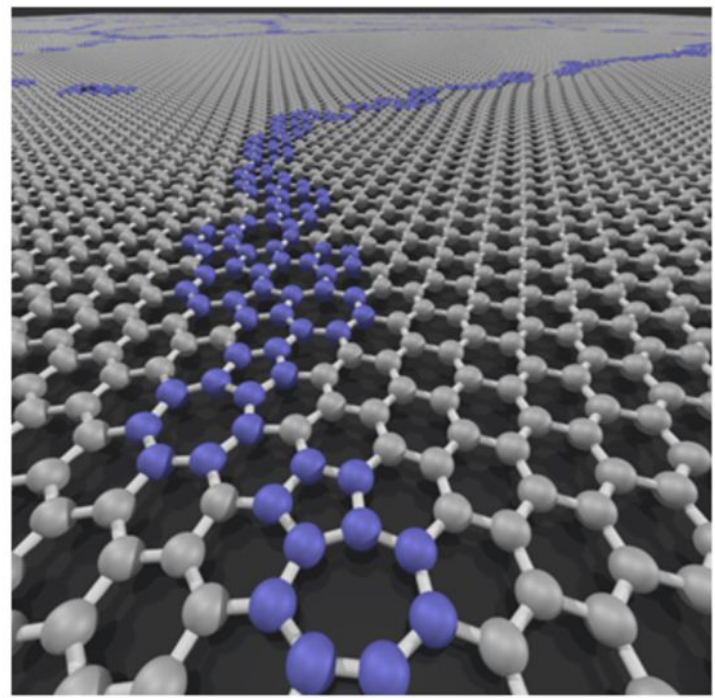

Figure 1

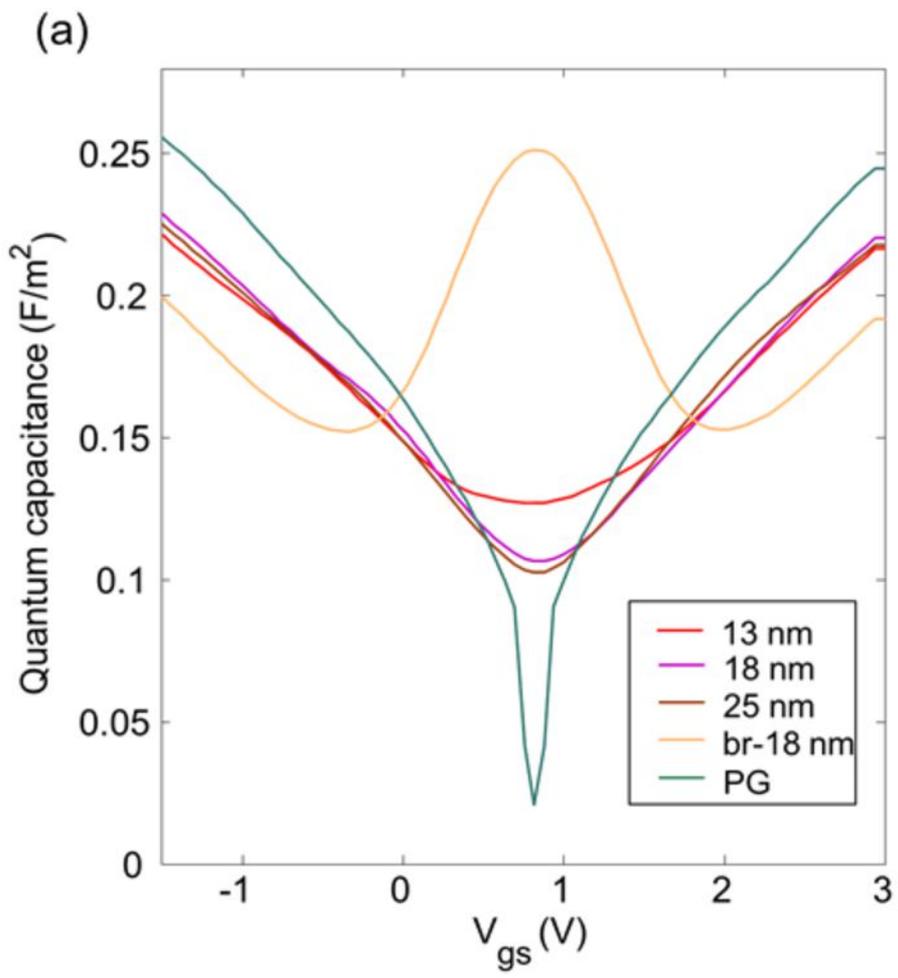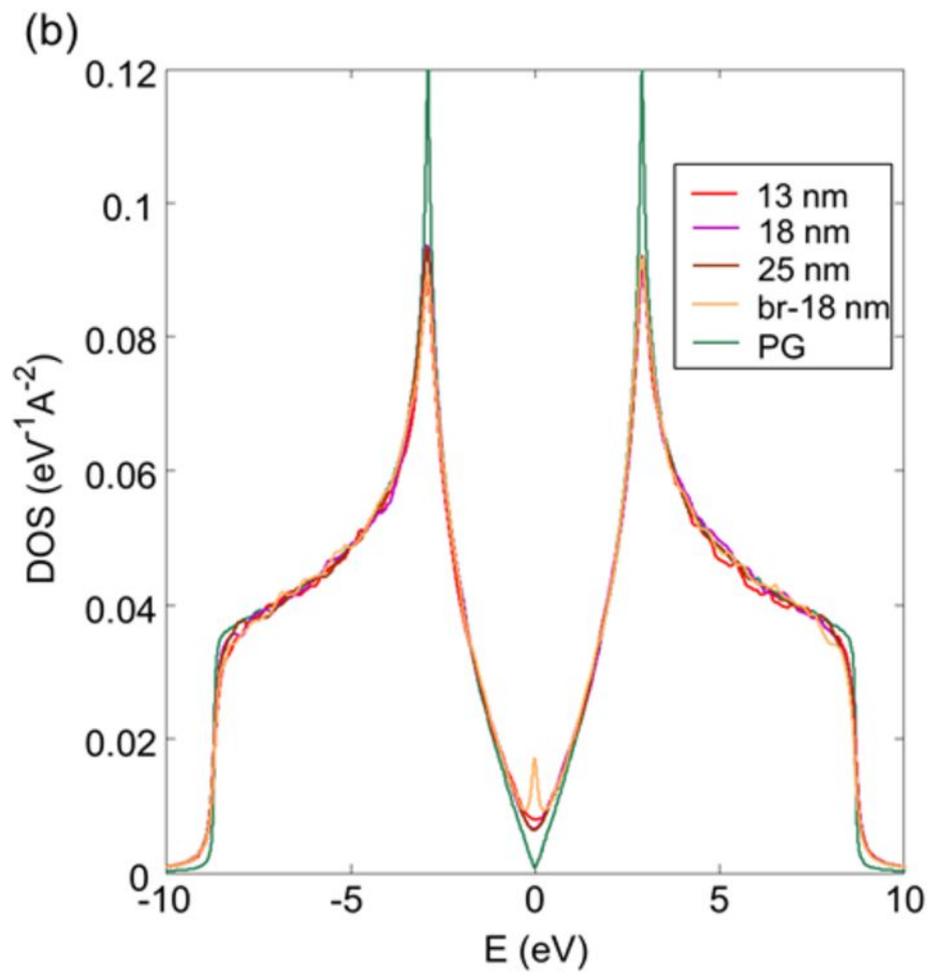

Figure 2

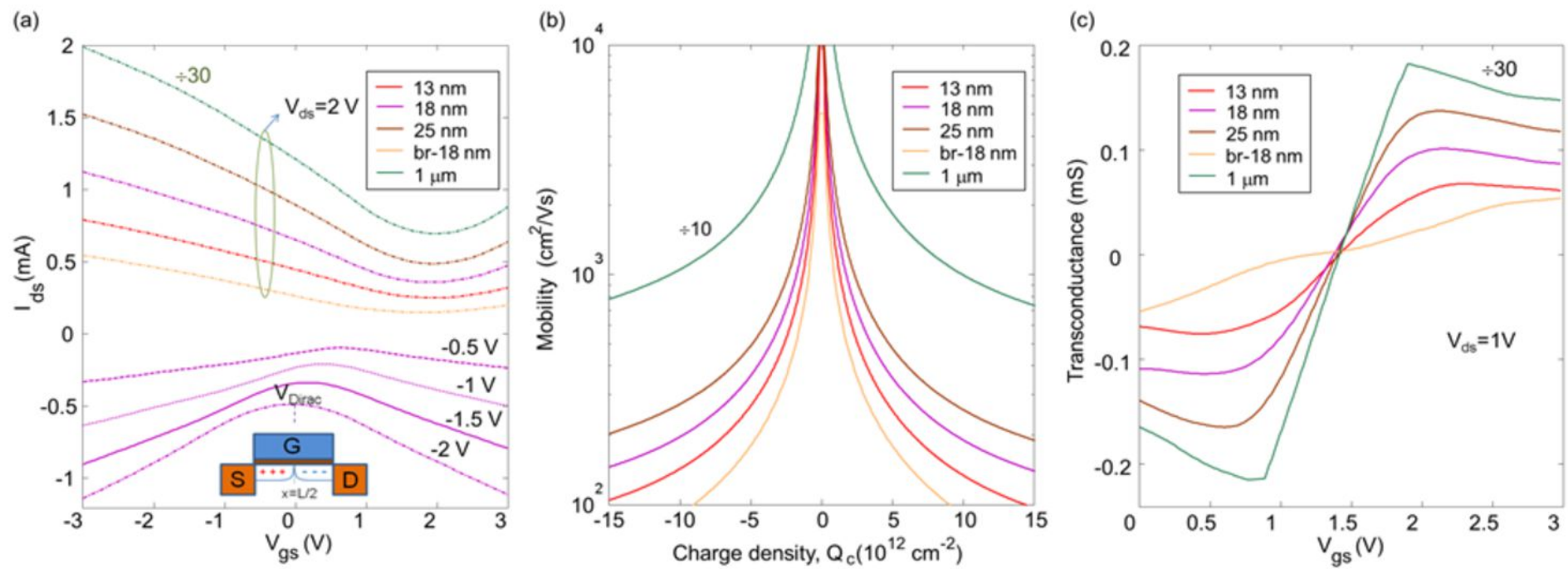

Figure 3

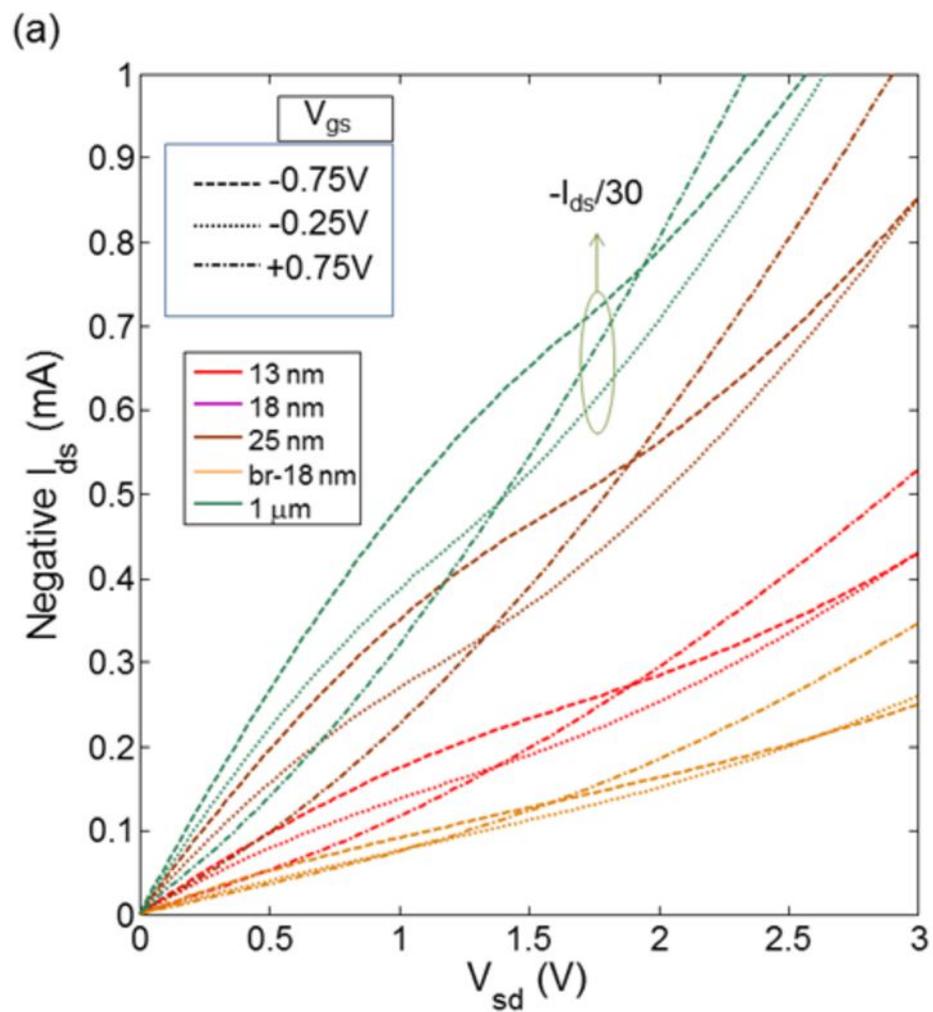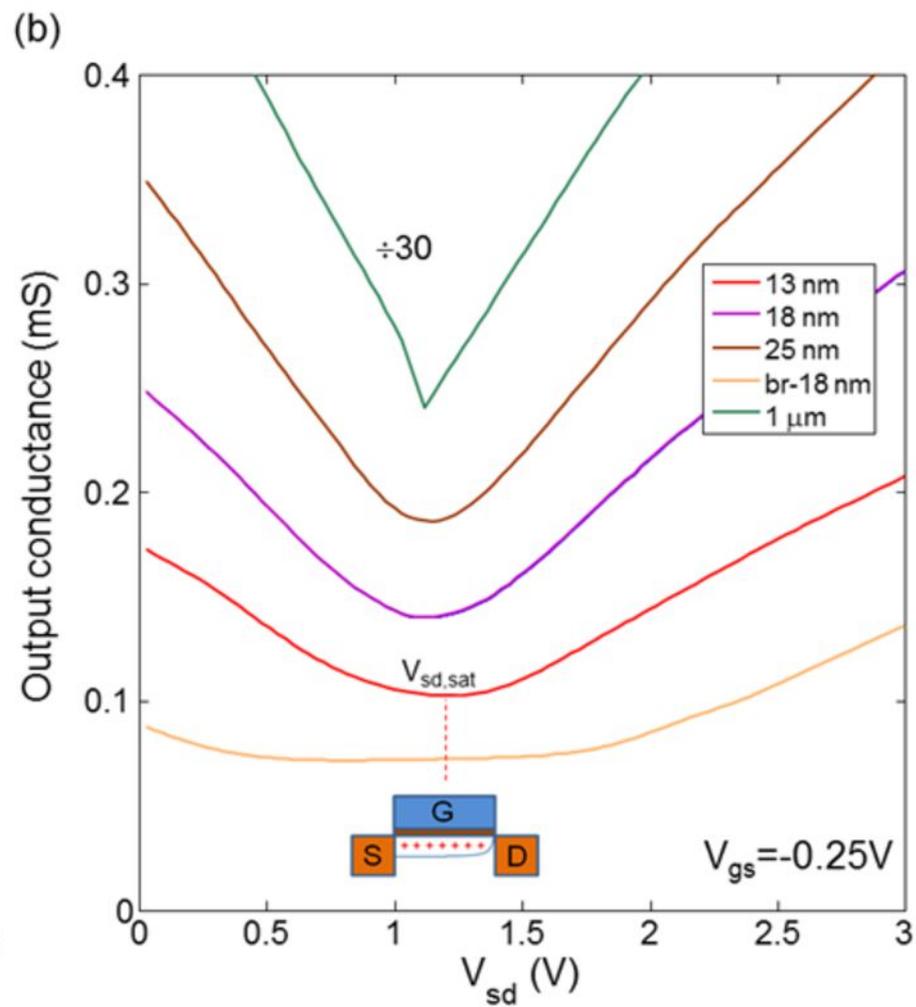

Figure 4

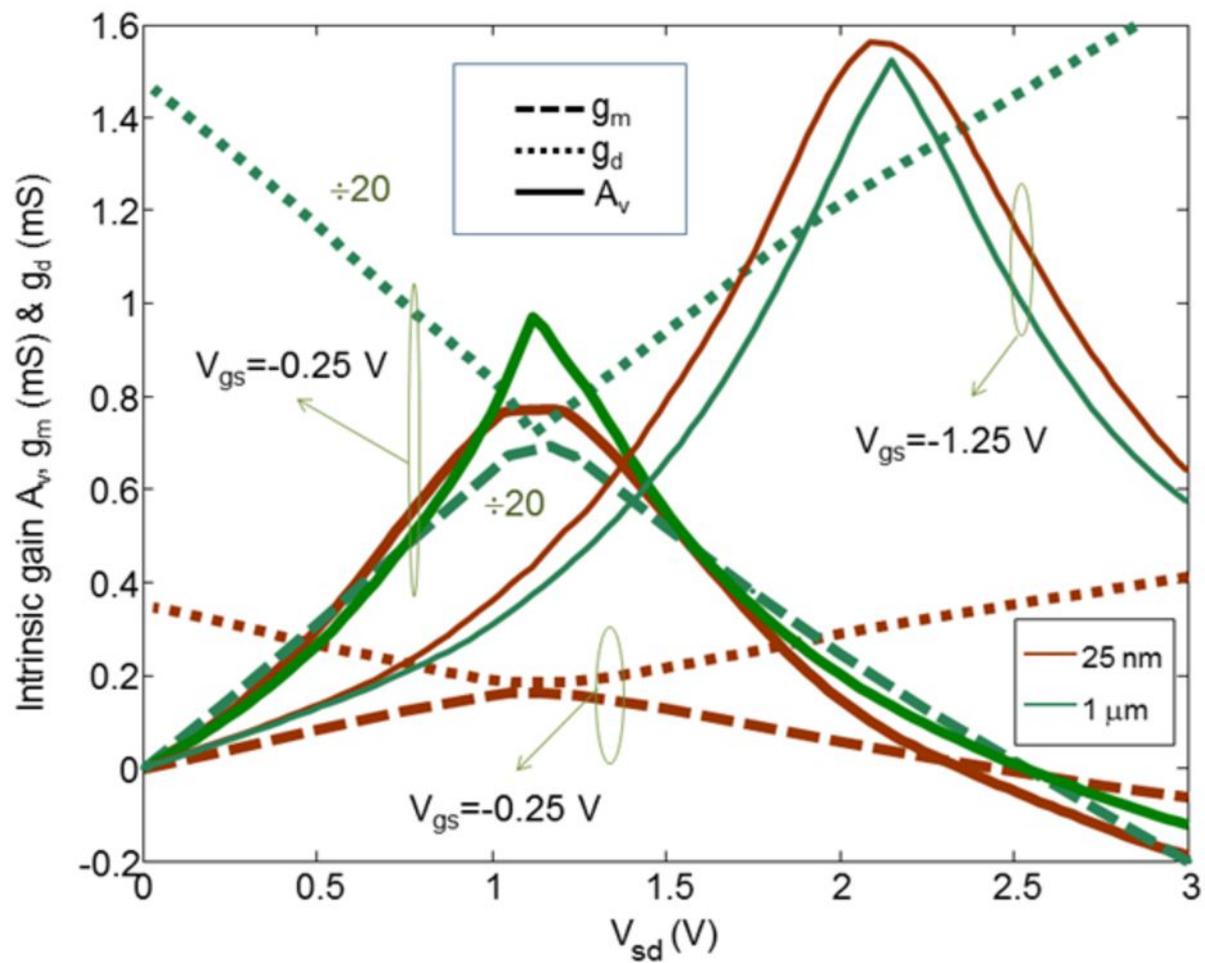

Figure 5

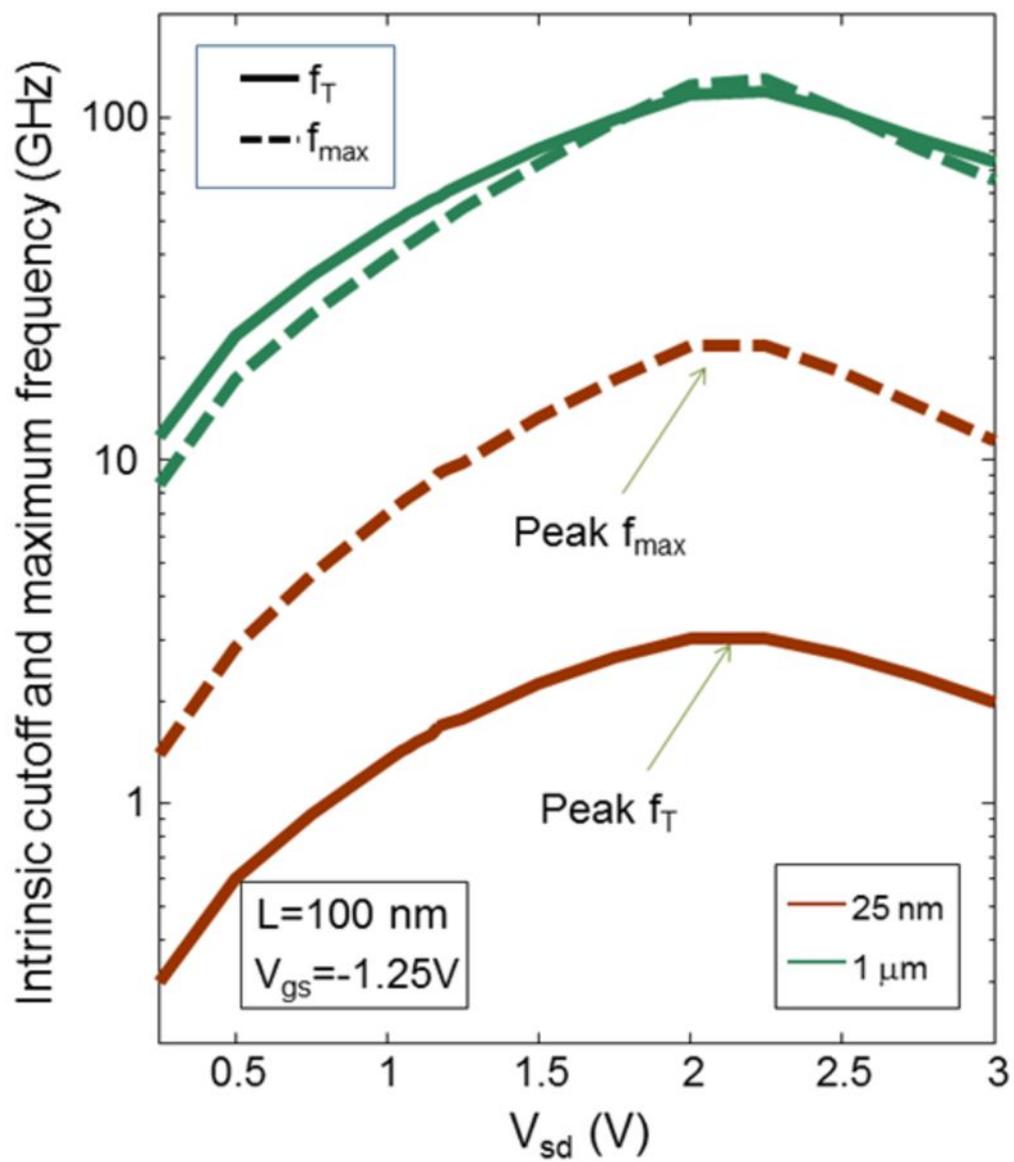

Figure 6